\documentclass[11pt,aps,prl,onecolumn,floatfix,longbibliography,a4paper,superscriptaddress]{revtex4-2}

\usepackage{amsmath, amssymb}
\usepackage{mathtools}
\usepackage{lipsum}
\usepackage{times}
\usepackage{graphicx}
\usepackage{wasysym}
\usepackage{bm}
\usepackage{hyperref}
\usepackage{xspace}
\usepackage{siunitx}
\usepackage{color}
\usepackage{braket}
\usepackage{todonotes}
\presetkeys{todonotes}{size=\footnotesize, color=yellow!90}{}

\makeatletter
\def\@fnsymbol#1{\ensuremath{\ifcase#1\or *\or \dagger\or *\or
   \mathsection\or \mathparagraph\or \|\or **\or \dagger\dagger
   \or \ddagger\ddagger \else\@ctrerr\fi}}
\makeatother

\definecolor{myColor}{rgb}{0.02,0.12,0.3}
\definecolor{myciteColor}{rgb}{0.39,0.7,0.89}
\hypersetup{colorlinks=true, linkcolor=myColor, filecolor=myColor, urlcolor=myColor, citecolor=myColor, urlcolor=myColor}
\DeclareSIUnit\angstrom{\text {Å}}

\DeclareSIUnit{\nK}{\nano\kelvin}
\DeclareSIUnit{\aB}{\emph{a}_0}
\DeclareSIUnit{\G}{G}

\renewcommand{\figurename}[1]{Fig.~}
\renewcommand{\thefigure}{\arabic{figure}}

\newcommand{\kB}{k_{\text{B}}}

\newcommand{\g}{\tilde{g}}

\newcommand{\Tc}{T_{\text{c}}}
\newcommand{\Nc}{N_{\text{c}}}

\newcommand{\G}{\Gamma_{\text{coll}}}

\def\ignore#1{}

\begin{document}


\title{Compressibility and the Equation of State of an Optical Quantum Gas in a Box}
\author{Erik Busley}
\author{Leon Espert Miranda}
\author{Andreas Redmann}
\author{Christian Kurtscheid}
\author{Kirankumar Karkihalli Umesh}
\author{Frank Vewinger}
\author{Martin Weitz}
\author{Julian Schmitt}
\email{schmitt@iap.uni-bonn.de}
\affiliation{Institut für Angewandte Physik, Universität Bonn, Wegelerstraße 8, 53115 Bonn, Germany}
\date{\today}

\begin{abstract}
The compressibility of a medium, quantifying its response to mechanical perturbations, is a fundamental property determined by the equation of state. For gases of material particles, studies of the mechanical response are well established, in fields from classical thermodynamics to cold atomic quantum gases. Here we demonstrate a measurement of the compressibility of a two-dimensional quantum gas of light in a box potential and obtain the equation of state for the optical medium. The experiment is carried out in a nanostructured dye-filled optical microcavity. We observe signatures of Bose-Einstein condensation at high phase-space densities in the finite-size system. Strikingly, upon entering the quantum degenerate regime, the measured density response to an external force sharply increases, hinting at the peculiar prediction of an infinite compressibility of the deeply degenerate Bose gas.
\end{abstract}
\maketitle
\thispagestyle{plain}

\pagestyle{plain}

Quantum gases of atoms, exciton-polaritons, and photons provide a test bed for many-body physics under both in- and out-of-equilibrium settings~\cite{Bloch:2008,Diehl:2008,Carusotto:2013}. Experimental control over dimensionality, potential energy landscapes, or the coupling to reservoirs offer wide possibilities to explore different phases of matter. For cold atomic gases, thermodynamic susceptibilities and transport properties have been extracted from density measurements~\cite{Navon:2010,Ho:2010a,Yefsah:2011,Hung:2011,Ku:2012,Mordini:2020} and have proven to be direct manifestations of the equation of state (EOS). In general, the EOS of a material, {\itshape e.g.}, its pressure–volume relation, describes both the thermodynamic state of a system under a given set of physical conditions as well as its response to perturbations, as mechanical compression. Experimental investigations of the EOS in quantum gases constitute a tool for the characterisation of phases and the identification of phase transitions, enabling important tests of physical models in a wide range of systems, from the ideal gas to superfluids and the interior of stars. 

Quantum gases of light have so far been experimentally realized in low-dimensional settings, mostly two-dimensional (2D) systems~\cite{Carusotto:2013}. Thermalized photon gases with non-vanishing chemical potential $\mu$, as well as Bose-Einstein condensation (BEC) have been demonstrated in dye-filled optical microcavities at harmonic confinement~\cite{Klaers:2010,Marelic:2015,Greveling:2018}, including measurements of density-insensitive thermodynamic quantities~\cite{Damm:2016}. In contrast, the isothermal compressibility $\kappa_T = n^{-2} (\partial n/\partial \mu)_T$ at temperature $T$ depends on the (local) particle density $n$ in the gas; for a systematic study, it thus is desirable to avoid spatially inhomogeneous density distributions inherent to harmonically trapped gases, and instead prepare uniform samples, where applying a spatially uniform force directly allows one to compress the gas and probe $\kappa_T$.

Notably, BEC does not occur in the infinite 2D homogeneous Bose gas given that thermal fluctuations at finite temperatures destroy long-range order~\cite{Mermin:1966}. While interactions nevertheless stabilize a superfluid through the Berezinskii-Kosterlitz-Thouless (BKT) transition, the infinite 2D ideal gas is doomed to remain quantum degenerate without forming a condensate. For a finite-sized homogeneous gas in a box, however, condensation is expected to be possible if the correlation length exceeds the system size at large phase-space densities~\cite{Hadzibabic:2011}. In ultracold atoms, the crossover between saturation-driven BEC and interaction-driven BKT superfluidity has been investigated in 2D harmonically trapped Bose gases by tuning the interactions exploiting a Feshbach resonance~\cite{Fletcher:2015}, while studies of homogeneous gases in box potentials have focused on the interacting regime~\cite{Ville:2018,Bohlen:2020,Christodoulou:2021}. In uniform gases of exciton-polaritons~\cite{Estrecho:2021}, on the other hand, the observation of BEC is hampered by reservoir-induced interactions and non-equilibrium effects. Importantly, up to now the compressibility and the EOS have not been determined for optical quantum gases.

Here we examine a 2D quantum gas of photons in a box potential. In the finite-size homogeneous system, we observe BEC, as evidenced from the measured position and momentum distributions. In subsequent experiments, a mechanical force is exerted onto the photon gas prepared in a regime around the phase transition. By studying the density response to minute forces, we measure both the bulk isothermal compressibility and the EOS of the optical quantum gas.

\begin{figure}[t] 
  \includegraphics{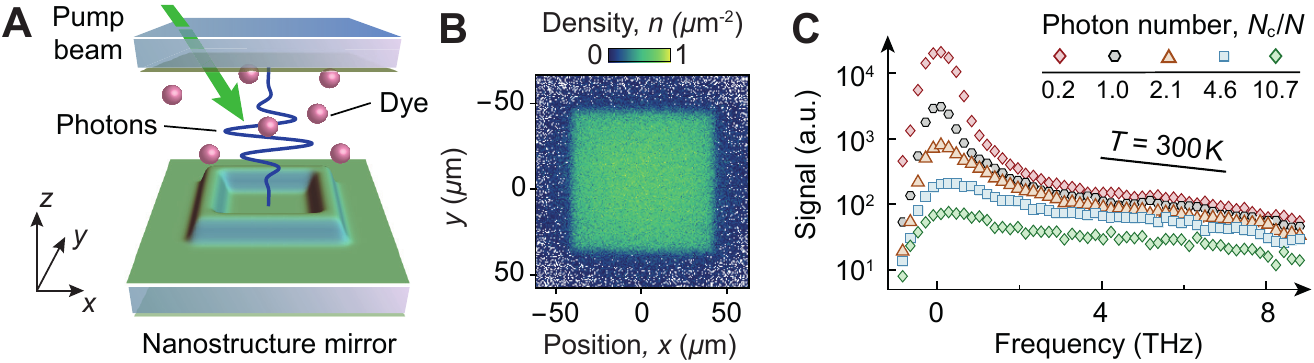}
 \caption{{\bf Trapping a homogeneous 2D photon gas.} ({\bf A}) A spatially structured microcavity filled with dye molecules confines and thermalizes the uniform photon gas. The locally elevated surface of one of the cavity mirrors realizes a repulsive box potential. ({\bf B}) Example surface density of a uniform photon gas in a square box with $L=\SI{80}{\micro\meter}$. ({\bf C}) Spectral distributions of the cavity emission exhibit an exponential decay consistent with $T\approx \SI{300}{\kelvin}$, indicating a thermal equilibrium photon gas. The population in the low-frequency modes is enhanced as $\Nc/N$ approaches unity.
}
\label{fig:1}
\end{figure}

Our homogeneous 2D photon gases are prepared in a nanostructured optical microcavity filled with a liquid dye solution~\cite{SI}, as illustrated in Fig.~\ref{fig:1}(A). The photons in the short cavity with mirror spacing on the order of the optical wavelength form a 2D gas of particles with an effective mass $m$, which are described by their transverse momentum $k=\sqrt{k_x^2+k_y^2}$. To spatially confine the unbound plane-wave states of a 2D homogeneous system, we implement a box potential of size $L$ as a container for the photons. The box potential is realized using a nanostructuring technique~\cite{SI,Kurtscheid:2020}, which imprints a position-dependent static surface elevation onto one of the cavity mirrors. The locally reduced cavity length results in a repulsive potential. Thermalization of the photon gas to room temperature is achieved by absorption and re-emission processes on the dye molecules~\cite{Klaers:2010}. Figure~\ref{fig:1}(B) shows an exemplary density distribution of a trapped gas recorded by imaging the cavity emission, and Fig.~\ref{fig:1}(C) gives optical frequency spectra over the trapped range $V_0/\hbar\simeq 2\pi\times \SI{9}{\tera\hertz}$ above the cavity low-frequency cutoff $\nu_\mathrm{c}=m c^2/(2\pi\hbar)$, where $\hbar$ is the reduced Planck’s constant and $c$ the speed of light, for different total particle numbers $N$. All spectra show an exponential decay of the population in the high-frequency states consistent with $\kB/(2\pi\hbar)\times \SI{300}{\kelvin}$, with Boltzmann's constant $\kB$, which we attribute as evidence for the gas to be thermalized. Moreover, the population in the lowest-lying states is enhanced as we increase $N$ beyond a critical value $\Nc$, which signals the emergence of a low-entropy phase as we discuss in detail in the following.

\begin{figure}[t] 
  \centering
  \includegraphics{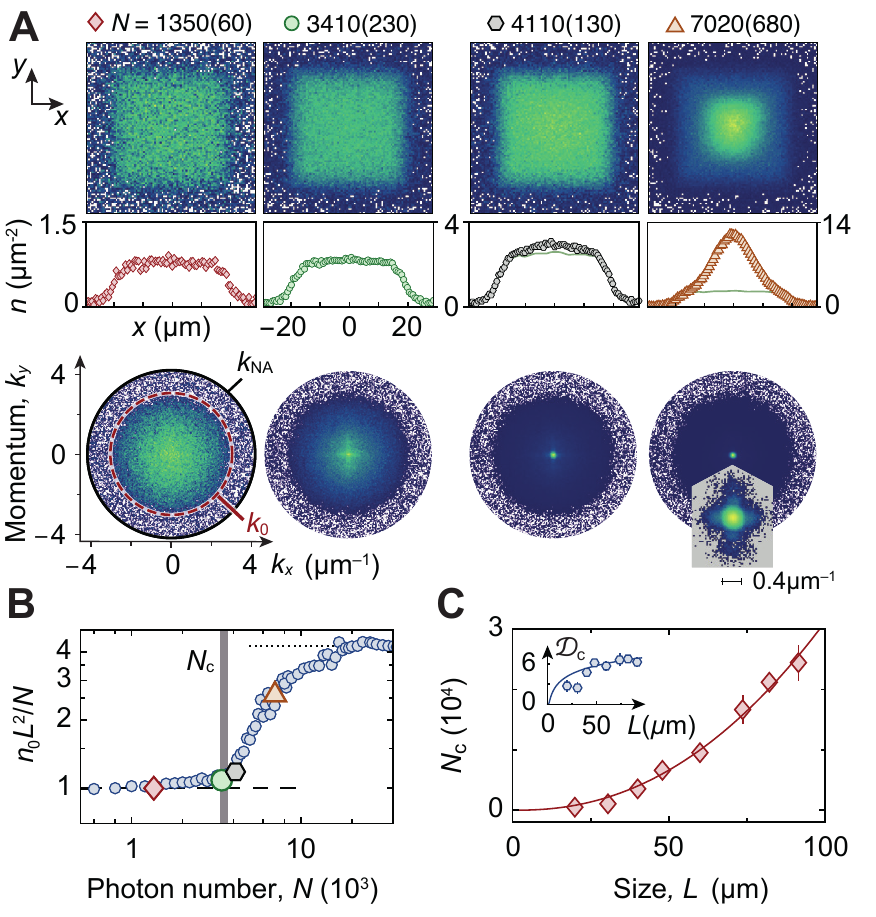}
 \caption{
 {\bf Bose-Einstein condensation of photons in a box.} ({\bf A}) Density distributions (top), line profiles (middle) and momentum-space distributions (bottom; $k_0$ denotes the trap limit, $k_\mathrm{NA}$ the imaging limit) as $N$ is increased beyond $\Nc=3.5(4)\times 10^3$. Below $\Nc$, the flat-top density and Gaussian $k$-space distribution resemble a thermal gas in the normal phase. Above $\Nc$, the ground state becomes massively populated and deforms the cloud, more directly observed in momentum space. ({\bf B}) Normalized central spatial density versus $N$, starting from $n_0 L^2/N=1$ in the normal phase, and approaching the expected value of $4$ when the ground state dominates (symbols as in (A)). ({\bf C}) $\Nc$ and $\mathcal{D}_\mathrm{c} =\Nc(\lambda/L)^2$ (inset), extracted from data as in (B), exhibit the predicted scaling (line) with the box size.
 }
\label{fig:2}
\end{figure}

We explore the quantum degenerate behavior in the finite-size homogeneous system. Figure~\ref{fig:2}(A) shows surface densities in the box for different $N$. Below the critical photon number $\Nc$, the bulk density is uniform as for a normal gas, while above, a macroscopic occupation of the ground state $|\psi_{1,1}(x,y)|^2=4L^{-2}\cos^2(\pi x/L)\cos^2(\pi y/L)$ with $|x|,|y|\leq L/2$ is observed; the qualitative change is evident from the line profiles. The corresponding momentum distributions below $\Nc$ in Fig.~\ref{fig:2}(A) resemble a Maxwell-Boltzmann distribution $f(k)\propto \exp(-k^2/\sigma_k^2)$ with $\sigma_k=\sqrt{2 m \kB T}/\hbar$; for our data, $\sigma_k=\SI{2.4(1)}{\micro\meter}^{-1}$ gives $T=\SI{295(33)}{\kelvin}$. Upon increasing the photon number, the population at small $k$ is enhanced and ultimately dominated by a strongly occupied ground state at $k=0$, the condensate. For large $k$, we observe signatures of both microscope aperture and finite trap depth~\cite{SI}. A closer inspection in $x$ and $k$-space (Fig.~\ref{fig:2}(A), inset) confirms that the ground state is Heisenberg-limited with an uncertainty product $\Delta x\Delta k=0.7(1)$, in agreement with theory $\Delta x\Delta k=\sqrt{\pi^2/12-1/2}\simeq 0.6$.

To quantify the transition point, we study the normalized spatial central density $n_0 L^2/N$ as a function of the particle number. Figure~\ref{fig:2}(B) shows the transition at $\Nc=3.5(4)\times 10^3$; the limiting cases $n_0 L^2/N=1$ ($=4$) for small (large) $N$ are well understood to arise from the uniform normal gas (inhomogeneous condensate) density with $N/L^2$ ($4N/L^2$) at the center. The critical particle number scales with the predicted $\Nc\sim(L/\lambda)^2 \log(L/\lambda)$, see Fig.~\ref{fig:2}(C), where $\lambda=2\sqrt{\pi}/\sigma_k\simeq \SI{1.47}{\micro\meter}$ denotes the thermal wavelength. The logarithmic scaling of the critical phase-space density $\mathcal{D}_\mathrm{c}$ shown in the inset is understood from the dependence of the coherence length reaching the system size~\cite{Hadzibabic:2011,SI}. At the largest investigated box sizes, we have $\mathcal{D}_\mathrm{c}=\Nc(\lambda/L)^2=6.3(8)$. For interacting 2D gases characterized by an interaction strength $\g$ and realized in ultracold atoms~\cite{Yefsah:2011,Hung:2011,Hadzibabic:2011,Fletcher:2015,Ville:2018,Christodoulou:2021}, the BKT phase transition to a superfluid usually occurs before BEC; for example, for homogeneous gases, $\mathcal{D}_\mathrm{c,BKT} =\log(380/\g)=6.5$ for $\g=0.6$~\cite{Christodoulou:2021}. Quite distinctly, for photon gases in dye-microcavities, self-interactions with $\g\leq 10^{-6}$~\cite{Dung:2017} imply a significantly larger $\mathcal{D}_\mathrm{c,BKT}\geq 20$ and, accordingly, both phases are expected to be well separated.

\begin{figure} [t] 
  \centering
  \includegraphics{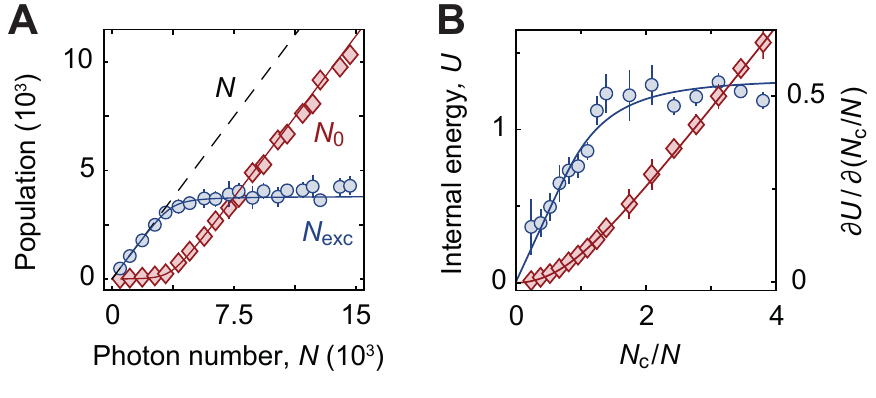}
\caption{
 {\bf Saturation and caloric properties.} ({\bf A}) Population in the ground (red) and excited states (blue), $N_0$ and $N_\mathrm{exc}$, respectively, extracted from $k$-space distributions (as in Fig.\ref{fig:2}(A), bottom) versus photon number $N$. At large $N$, the visible saturation of the normal part $N_\mathrm{exc}$ indicates that the phase transition is BEC-like. ({\bf B}) Normalized internal energy (red), showing a crossover from quadratic (condensed gas, $\Nc/N<1$) to linear (normal, $\Nc/N\gg 1$) scaling, and the derivative (blue), which reflects the specific heat per particle, versus $\Nc/N$. In the normal phase, the latter reaches a value near $0.5$, which is understood from the finite trap depth $V_0$, and is below the value of $1$ expected for $V_0\rightarrow\infty$. Solid lines are finite-size theory.
}
\label{fig:3}
\end{figure}

We identify the BEC-like nature of the phase transition by extracting the ground and excited state populations, $N_0$ and $N_\mathrm{exc}$, respectively, from the momentum space distributions as a function of the total particle number $N$. Figure~\ref{fig:3}(A) shows the visible saturation of the normal part, which indicates that interaction effects are very small~\cite{Tammuz:2011}; in particular, it gives evidence that in our homogeneous finite-size system BEC is prevalent, instead of BKT. This interpretation is supported by the deduced caloric properties of the gas, see Fig.~\ref{fig:3}(B), which closely follow the ideal Bose gas prediction. For the internal energy $U=\langle E\rangle \Nc/N^2 \kB T$, where $\langle E\rangle$ denotes the average transverse energy, we observe a crossover from a quadratic to a linear scaling in the condensed and normal phase, respectively, as a function of $\Nc/N$. Correspondingly, its derivative $\partial U/\partial(\Nc/N)$ is a smooth function, highlighting that the heat capacities in the normal and condensed phases are linked without any discontinuities~\cite{Ku:2012,Damm:2016}. In the normal-gas phase each particle can accommodate only $\approx 0.5 \kB T$ of thermal energy, as well understood from the finite trap depth~\cite{SI}.

\begin{figure} [t] 
  \centering
  \includegraphics{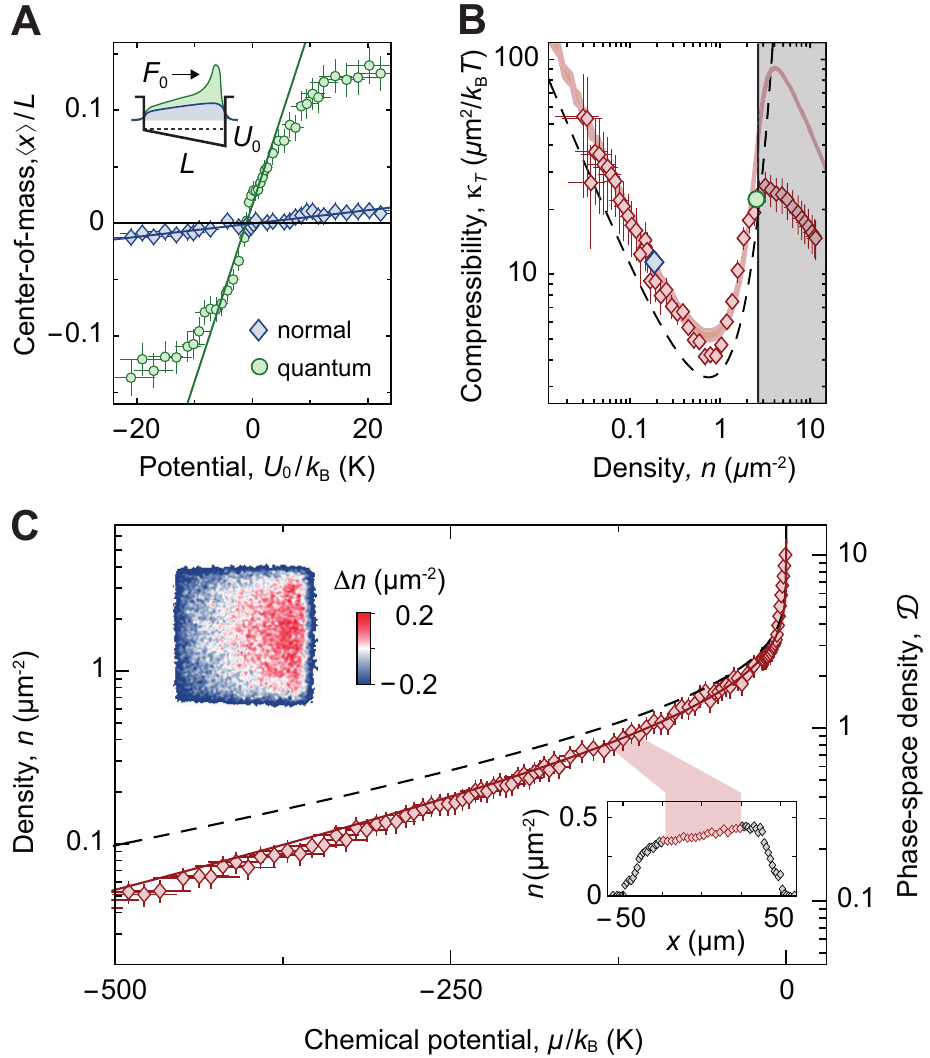}
\caption{
 {\bf Compressibility and equation of state.} ({\bf A}) A linear potential realizes a force $F_0=U_0/L$ that compresses the photon gas (inset). Depending on its normal (blue) or quantum degenerate (green) state, the gas shows a different center-of-mass response, fitted in the linear response region (lines). ({\bf B}) The compressibility $\kappa_T$ decreases with density $n=N/L^2$, until above $n\simeq\SI{0.8}{\micro\meter}^{-2}$ quantum degeneracy sets in and the Bose gas becomes highly compressible as a consequence of the gradually vanishing energy cost to increase the density. At $n\geq \SI{2.6}{\micro\meter}^{-2}$ (shaded), the $\kappa_T$ extraction ceases to be valid; this range of validity has been confirmed numerically~\cite{SI}. For comparison, we show finite-size theory (red line, shading indicates uncertainty in $V_0$, which causes the vertical shift), where the kink at large densities is due to the finite box size, and the prediction for the infinite 2D Bose gas (dashed). ({\bf C}) From the density profiles (insets: exemplary density deviation from $U_0=0$ and line profile) we extract the equation of state by combining data for different local chemical potentials~\cite{SI}. Lines show theory for finite-depth box (solid) and infinite-system (dashed).
 }
\label{fig:4}
\end{figure}

To determine the isothermal compressibility $\kappa_T = n^{-2} (\partial n/\partial\mu)_T$ of the optical quantum gas, we exert a force onto the photons by tilting one of the cavity mirrors, which superimposes a linear potential $U(x)=U_0 x/L$ to the box, and measure the density response. Figure~\ref{fig:4}(A) shows the displaced center-of-mass $\langle x \rangle$ as a function of the tilt $U_0$. In local density approximation (LDA), with chemical potential $\mu(x)=\mu_0-U(x)$, the center-of-mass to first order relates to the compressibility, following $\langle x\rangle/L = -\kappa_T n U_0/12$~\cite{Ho:2010a,SI,Pitaevskii:2016}. For small $U_0$, the data in Fig.~\ref{fig:4}(A) confirms the linear behavior and shows an enhancement of the density response when going from the normal to the condensed phase. The visible saturation for large $U_0$ is caused by the finite box size, which limits the displacement.

Figure~\ref{fig:4}(B) shows the compressibility, extracted from a linear fit of the region near $U_0=0$, see Fig.~\ref{fig:4}(A), along with theory for the infinite and finite system. Below the critical density for condensation $\Nc/L^2\simeq \SI{2.6}{\micro\meter}^{-2}$, the photon distribution is spatially homogeneous, as visible in Figs.~\ref{fig:2}(A,B). Here LDA can be applied~\cite{Pitaevskii:2016} to extract $\kappa_T$ using data with small tilts ($U_0\lesssim\mu$), even in the case of very small interactions~\cite{SI}. In the condensed phase, on the other hand, the LDA ceases to be valid, and the corresponding region is shaded in gray. Within the region of validity, the compressibility compares well with theory. Remarkably, at densities $n\geq \SI{1}{\micro\meter}^{-2}$, we observe a sharp increase of the compressibility; the onset is in good agreement with the prediction for an infinite non-interacting system (dashed line)~\cite{SI}. The corresponding function $\kappa_T = [\exp(n \lambda^2)-1]/(\kB T n^2 \lambda^2)$ exhibits a minimum at $n\simeq 1.59/\lambda^2\simeq \SI{0.74}{\micro\meter}^{-2}$, close to the measured value of $\SI{0.79(5)}{\micro\meter}^{-2}$. It is well understood that as the thermal wave packets spatially overlap the classically expected decrease in compressibility with density (it is harder to compress a dense gas than a dilute one) is replaced by a compressibility increase stemming from the quantum-statistical occupation of low-lying energy levels, reducing the energy cost for compression as compared to the classical gas case. In the extreme high-density limit of an infinite-size deeply degenerate gas, bosons can be added to the system at essentially vanishing energy cost, meaning that $(\partial\mu/\partial n)_T$ gradually approaches zero as $\mu\rightarrow 0$, so that the compressibility takes arbitrarily large values.

Finally, we study the equation of state $n=f(\mu,T)$ of the photon gas. Figure~\ref{fig:4}(C) shows the variation of the density $n$ as a function of the chemical potential $\mu$, as determined from combining recorded density profiles of the gas in a tilted box at different $N$; the insets give an exemplary density deviation from the unperturbed case and the corresponding line profile $n(x)$. As above, the conversion from position to chemical potential relies on the potential gradient $U_0/L\approx \kB\  \SI{0.6}{\kelvin/\micro\meter}$, which acts as a calibrated differential energy scale $\mathrm{d}\mu=U_0/L\thinspace \mathrm{d}x$. Note that our data exhibits a larger slope than the EOS of the infinite 2D Bose gas, $n(\mu)= -\lambda^{-2}\log[1-\exp(\mu/\kB T)]$ (dashed line) owing to the finite trap depth, as we confirm by numerical calculations~\cite{SI}. Except for the condensed regime, where LDA is invalid, our method reliably extracts the EOS of a quantum gas of light. 

We have demonstrated a measurement of the compressibility of an optical quantum gas and determined its EOS. The experiment is carried out using a 2D photon gas inside a box potential, both below and above the phase transition to a BEC. Compression of optical gases may have direct consequences for thermodynamic machines with light as a work medium~\cite{Ghosh:2019}. A further interesting perspective is the exploration of sound~\cite{Ville:2018,Bohlen:2020,Christodoulou:2021,Estrecho:2021}. The required dynamic manipulation of optical quantum gases is feasible by, {\itshape e.g.}, electro-optic trap modulation or spatiotemporally resolved pumping of the dye reservoir~\cite{Schmitt:2015}. Beyond ideal gas theory, a nonvanishing healing length can be achieved by adding either Kerr media or exploiting the weakly dissipative nature of photon condensates~\cite{Gladilin:2020}. The demonstrated homogeneous quantum gas of light in a box opens new possibilities for studies of universal phenomena in 2D, including critical behavior~\cite{Comaron:2018} and the non-equilibrium Kardar-Parisi-Zhang phase~\cite{Zamora:2017}.
 
%
%
%
%
\vspace{0.5cm}
\noindent\textbf{Acknowledgments.} We thank F. König for experimental assistance; and D. Luitz, P. Christodoulou, Z. Hadzibabic, and R. Lopes for fruitful discussions. This work was supported by the DFG within SFB/TR 185 (277625399) and the Cluster of Excellence ML4Q (EXC 2004/1–390534769), and by the EU within the Quantum Flagship project PhoQuS (820392). J.S. acknowledges support from an ML4Q Independence grant. 

\bibliography{references}


\setcounter{figure}{0} 
\setcounter{equation}{0} 
\renewcommand\theequation{S\arabic{equation}} 
\renewcommand\thefigure{S\arabic{figure}}

\section*{Supplementary Information}

\subsection{Experimental scheme}
\begin{figure}[t] 
  \centering
  \includegraphics[width=0.6\columnwidth]{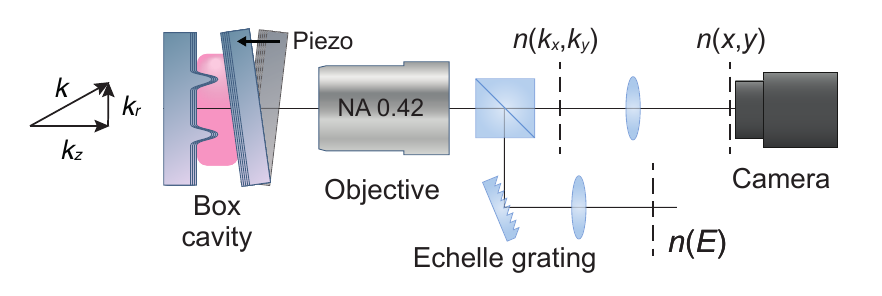}
 \caption{{\bf Sketch of the dye-filled microcavity experiment.} Photons transmitted through one of the cavity mirrors are collected by a microscope objective, and their position $n(x,y)$, momentum $n(k_x,k_y)$ or energy $n(E)$ spectral distributions are detected using intensified cameras. To precisely co-align the cavity mirrors (or induce linear gradient potentials) calibrated piezo mounts are used.
}
\label{fig:S1}
\end{figure}

The photon gas is operated inside a dye-filled optical microcavity, see Fig.~\ref{fig:S1}, which consists of two plane Bragg mirrors each with a maximum reflectivity of $99.998\%$ and spaced by $D_0=q\lambda_0/2\tilde n\approx \SI{1.4}{\micro\meter}$, where $q=7$ is the longitudinal wave number and $\lambda_0=\SI{583}{\nano\meter}$ the optical wavelength. As a dye medium we use rhodamine 6G solved in ethylene glycol (refractive index $\tilde n =1.44$). Owing to the small value of $D_0$, the wave vector component $k_z=\pi q/D_0\tilde n$ introduces a large energy gap $\hbar k_z c/q\gg \kB T$ between adjacent modes, with reduced Planck’s constant $\hbar$, the speed of light $c$, Boltzmann’s constant $\kB$ and temperature $T=\SI{300}{\kelvin}$. The gap effectively freezes out the motional degree of freedom of the photons along $z$, making the system 2D. In the transverse direction, the quasi-continuous momenta $k_{x,y}$ describe the in-plane motion of the particles in the homogeneous system. Our experiment allows the single-pulse-resolved detection of the cavity emission, and the acquired data is averaged over 5-10 shots. 

One of the cavity mirrors exhibits a reflectivity maximum at $\SI{550}{\nano\meter}$ and a bandwidth of $\SI{50}{\nano\meter}$, where the reflectivity is above $99.99\%$; in addition, it is equipped with an absorptive silicon layer below the high-reflectivity Bragg coating, which enables nanostructuring of the mirror surface (see section “Potential creation” and ref.~\cite{Kurtscheid:2020}). The opposing mirror consists of a custom dielectric coating, which contains two stacked reflection bands of $\SI{50}{\nano\meter}$ spectral width, the first one centered at $\SI{570}{\nano\meter}$ (on top; in contact with the dye), the second one around $\SI{700}{\nano\meter}$ (below; sandwiched between top-layers and glass substrate). While the top coating confines the photon gas, the lower coating prevents residual isotropic fluorescence from reaching the analysis part. The cavity emission transmitted through this mirror is guided into the analysis part of the experiment.

To maintain a steady-state photon number inside the cavity, continuous pumping is required to compensate losses from mirror transmission. We optically pump the dye with a laser at $\SI{532}{\nano\meter}$ under a $42^\circ$ angle (mirror transmission $>50\%$); the pump light is temporally chopped into $\SI{500}{\nano\second}$-pulses at $\SI{30}{\hertz}$ repetition rate to minimize photobleaching. Operating in the optically dense regime $\Gamma \tau\gg 1$, most fluorescence photons are reabsorbed by the dye film at a rate $\Gamma$ during the cavity lifetime $\tau$. This leads to a detailed balance between molecules and photons if the ratio of absorption and emission rates scales with a Boltzmann-factor, which is the case provided the used dye fulfills the Kennard-Stepanov relation~\cite{Stepanov:1971}. The photon gas thermalizes with the dye heat bath and acquires the rovibrational temperature of the molecules, which is at room temperature~\cite{Klaers:2010}. At fixed $T$, the phase-space density of the photon gas is determined by the pumping; controlling the excitation level of the dye molecules allows us to vary the chemical potential of the photons and prepare gases in the normal to quantum-degenerate regime~\cite{Klaers:2012,Schmitt:2018}. To illuminate the system as homogeneously as possible, the pump beam diameter is adjusted individually to roughly twice the system size for each box.

The cavity emission is collected by a long working distance microscope objective (Mitutoyo $20\times$, $\mathrm{NA}=0.42$) and imaged onto an EMCCD camera (Andor iXon Ultra 897) to record the spatial- or momentum-space distributions, respectively, as sketched in Fig.~\ref{fig:S1}. The latter is achieved by two interchangeable sets of lenses in the optical path, enabling us to image either the cavity or the Fourier plane onto the camera sensor. The obtained momentum distributions shown in the bottom panels of Fig.~\ref{fig:2}(A) of the main text are cropped at the maximum transverse momentum $k_\mathrm{NA}$ collected by our microscope and allow to identify the trap depth $V_0=1.4 \kB T$, which sets an upper limit for the energy of the photons that can be contained in the box; for momenta beyond $k_0=\sqrt{2 m V_0}/\hbar\approx \SI{2.9}{\micro\meter}^{-1}$, the signal indeed vanishes. Moreover, we obtain broadband energy-resolved spectra of the cavity emission by dispersing the collimated light with a diffraction grating (1200 ll/mm) and detecting its first order on an ICCD camera; the spectral resolution is $\Delta\nu=\SI{0.6}{\tera\hertz}$ (FWHM), see Fig.~\ref{fig:1}(C) of the main text. To identify the individual energy levels of the photon gas in the box potential (Fig.~\ref{fig:S2}(E)), an Echelle grating with 316 ll/mm is at our disposal; operating in a high diffraction order gives a spectral resolution $\Delta\nu=\SI{19}{\giga\hertz}$ (FWHM).

To prepare uniform photon gases in the box and apply potential gradients for the compression of the optical gas by tilting the cavity mirrors, a high-precision alignment of the cavity mirrors is required. For example, a cavity length variation of $\SI{1}{\angstrom}$ at $q=7$ results in a potential energy $\hbar(2\pi\times 36)$~GHz, which exceeds the ground state energy $E_{1,1}=\hbar(2\pi\times 13)$~GHz in a $\SI{40}{\micro\meter}$ sized box, perturbing the ground state wave function. To achieve the desired sub-\AA\  precision, both cavity mirrors are fixed in 2" mirror mounts with complementary piezo actuators technologies: the first mount (Thorlabs POLARIS-K2S2P) is equipped with two continuous-voltage piezo actuators, which allow for step-free adjustment of the mirror tilt angle in $x$ and $y$ direction; the second mount uses piezo actuators (PI N-470, $\SI{20}{\nano\meter}$ step size), which allow for discrete, but well-countable calibrated mirror tilts. One step induces a height difference of 10pm over a $\SI{40}{\micro\meter}$ sized box, which corresponds to a potential energy difference $\kB \times \SI{0.16}{\kelvin}=\hbar(2\pi\times 3.3)$~GHz and is sufficient for alignment. For the compressibility measurements, we have verified that the discrete steps provide reproducible tilt angles.

\subsection{Potential creation}
\begin{figure}[t] 
  \centering
  \includegraphics{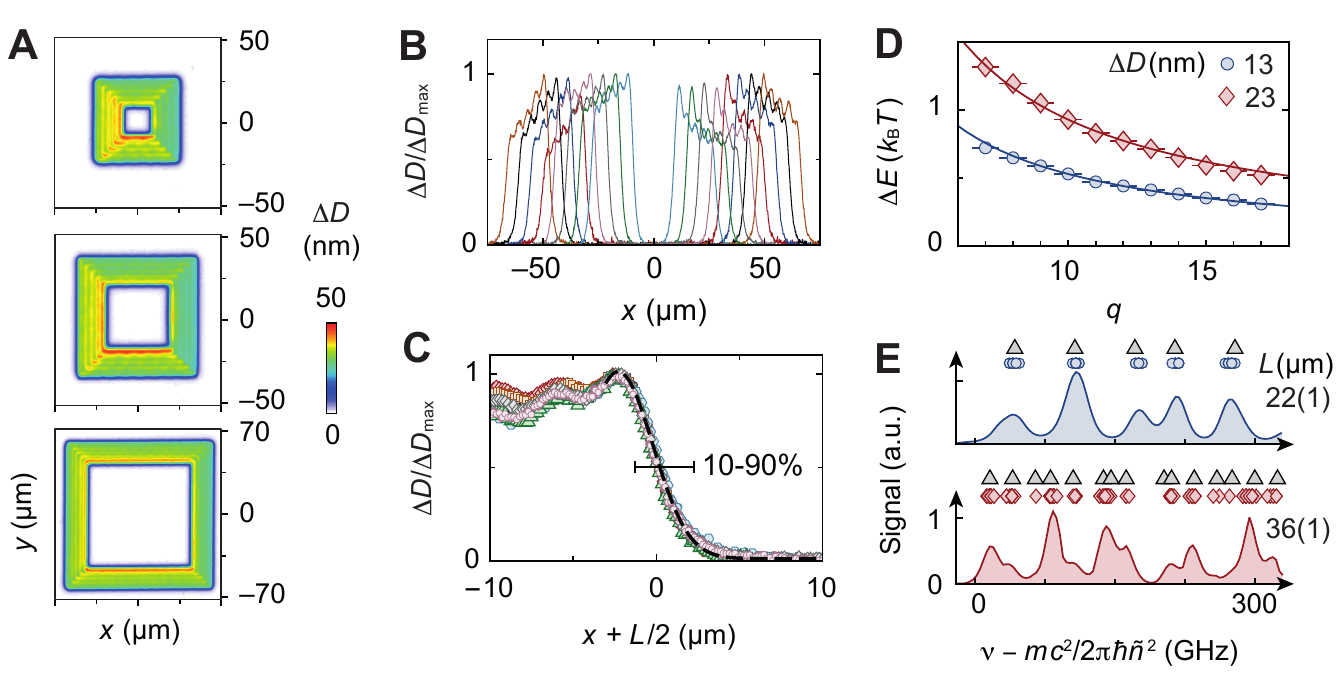}
 \caption{{\bf Nanostructure box potentials.} ({\bf A}) Height profiles of the nanostructured mirror surface for box potentials with $L\approx\{20,40,90\}\SI{}{\micro\meter}$. ({\bf B}) Normalized cuts of the mirror surface height $\Delta D$ for all investigated sizes $L\approx\{20,30,...,90\}\SI{}{\micro\meter}$. ({\bf C}) Zoomed-in view of the box slope after shifting the cuts from (B) horizontally; independent of $L$, all curves exhibit a similar decay from 90 to 10\% over $\SI{3.7(1)}{\micro\meter}$. ({\bf D}) The spectral width of the trapped photon gas emission versus $q$ for two nanostructures along with parameter-free theory based on eq.~\eqref{eq:S1} (lines). ({\bf E}) Lowest-lying energies of a particle in a box of size $L=\SI{22(1)}{\micro\meter}$ (blue, top panel) and $\SI{36(1)}{\micro\meter}$ (red, bottom), respectively, relative to the cavity low-frequency cutoff. The high-resolution spectra ($\Delta\nu=\SI{19}{\giga\hertz}$) show resonances (circles and diamonds, extracted from multiple spectra) close to the theory predictions (triangles).
}
\label{fig:S2}
\end{figure}

The box potentials are created by nanostructuring one of the plane mirrors prior to its operation in the dye-filled cavity, see Fig.~\ref{fig:S2} and ref.~\cite{Kurtscheid:2020} for a detailed description of the method. We note that recent other work has reported an alternative technique to create potentials for microcavity photon gases, including box potentials, using focused ion beam milling~\cite{Walker:2021}. In our work, the mirrors used as writing samples contain a $\SI{30}{\nano\meter}$-thin silicon layer below their dielectric Bragg coating, which enables the structuring process: A $\SI{532}{\nano\meter}$ laser beam is focused through the quartz glass substrate onto the silicon layer (beam diameter $\SI{1}{\micro\meter}$), where part of the light is absorbed and converted into heat. Above a threshold laser power near $\SI{30}{\milli\watt}$, the heating induces a local elevation of the dielectric layer stack with Gaussian transverse profile of $\SI{5}{\micro\meter}$ width (FWHM). The process is attributed to a thermally-induced mechanical stress enhancement between the alternating Bragg layer materials, which– above a threshold value– leads to a detachment, often called delamination, of neighboring layers~\cite{Goia:1997}. The 2D box structures shown in Fig.~\ref{fig:S2} are created by steering the focused laser beam across the silicon layer using a galvo scanner. We use unity-aspect-ratio box potentials of sizes $L=\{20,30,...,90\}\SI{}{\micro\meter}$ with a potential wall increasing from 10\% to 90\% over $\SI{3.7(1)}{\micro\meter}$ distance, independent of $L$. The achieved variable structure heights $\Delta D\leq \SI{60}{\nano\meter}$ are controlled by the laser power. For $\Delta D\geq \SI{25}{\nano\meter}$, however, we observe that transverse wave vectors $k_0 \geq \SI{3}{\micro\meter}^{-1}$ are not confined by our box; in other words, the spectra show a high-energy cutoff. The underlying mechanism of the effect is presently not understood and requires further studies. To mitigate this limitation, we restrict our studies in the present work to structure heights (trap depths) below $\SI{25}{\nano\meter}$ ($1.4 \kB T$), where the potential depth exhibits a systematic behavior in accordance with eq.~\eqref{eq:S1} (see section “Two-dimensional photon dispersion”).

The cavity mirrors (before nanostructuring) exhibit an r.m.s. surface roughness of $\delta D\approx \SI{0.4}{\angstrom}$ within a $(20\times 20)\SI{}{\micro\meter}^2$ region. For our experimental parameters, this corresponds to a 'noise floor' potential $\hbar(2\pi\times 15)\SI{}{\giga\hertz}$ of the same order as the ground state energy $E_{1,1}$ in a $\SI{40}{\micro\meter}$-sized box, and we observe its influence on the ground mode profiles in boxes with $L\geq\SI{50}{\micro\meter}$ due to the rapidly decreasing ground state energy, $E_{1,1}\propto 1/L^2$. For the used moderate-height nanostructures, we do not find evidence for a reduction of the mirror reflectivity; correspondingly, typical photon lifetimes $\tau\approx \SI{300}{\pico\second}$ in the box-cavity are sufficiently long to ensure thermalization of the photon gas, which is established within roughly the (spectrally averaged) absorption time $\Gamma^{-1}\approx \SI{50}{\pico\second}$ at our dye concentration of $1$~mmol/l and cutoff wavelength $\lambda_0=\SI{583}{\nano\meter}$.

\subsection{Two-dimensional photon dispersion}
Photons trapped in a microcavity environment exhibit a modified dispersion relation as compared to free photons. The highly anisotropic dimensions of the resonator ($D_0\approx\SI{1}{\micro\meter}$ longitudinally versus $L\approx 20-90\SI{}{\micro\meter}$ transversally) freeze out the motional degree of freedom in $z$ direction, making the photons formally equivalent to massive particles, which are deeply 2D and experience in-plane potentials imposed by the mirror surface height profiles. Starting from the free photon energy-momentum relation, $E=\hbar k c/\tilde n$ with wave vector $k=\sqrt{k_x^2+k_y^2+k_z^2}$, we apply the paraxial approximation $k_{x,y}\ll k_z=\pi q/(D_0-\Delta D)$ with local cavity length variation $\Delta D\equiv\Delta D(x,y)$; note that we define $\Delta D>0$ for delaminated, {\itshape i.e.}, elevated mirror surfaces. To first-order one obtains
\begin{equation}
E\approx m \left(\frac{c}{\tilde n}\right)^2 + \frac{\hbar^2(k_x^2+k_y^2)}{2m}+\left(\frac{c}{\tilde n}\right)^3 \frac{m^2}{\pi\hbar q}\Delta D,
\label{eq:S1}
\end{equation}
where $m=\pi\hbar q \tilde n/D_0 c$ is the effective photon mass, and the quadratic dependence on momentum underlines the phenomenology of a massive particle in 2D. The first term gives the rest energy of the photon, the second its kinetic energy, and the last term describes a potential energy, which is repulsive in our experiments with $\Delta D>0$. For typical experimental parameters, $q=7$ and $\lambda_0=\SI{583}{\nano\meter}$, we get $m=7.8\times 10^{-36}\SI{}{\kilogram}$ and a potential energy $\kB T$ every $\Delta D=\SI{17}{\nano\meter}$ of mirror surface elevation. To validate the trap depth $V_0$, we measure the energy width of the cavity emission spectrum $\Delta E$ for different resonator lengths $q$ and structure heights $\Delta D$. Figure~\ref{fig:S2}(D) shows the confirmed $1/q$-scaling of the potential depth predicted by eq.~\eqref{eq:S1} without any free parameters for two nanostructures. Note that experimentally $q \geq 7$ and $\Delta D < \SI{25}{\nano\meter}$, such that the achieved maximum trap depth is limited to about $1.4 \kB T$.

\subsection{Density, critical point, and thermodynamics}
We turn to the thermodynamics of the 2D photons in the box, derive their surface density distribution and define a critical particle number for BEC in the finite-size system. A particle in a square-box potential of infinite trap depth and size $L$, is restricted to quantized eigenenergies $E_{n_x,n_y}=\pi^2 \hbar^2 (n_x^2+n_y^2)/2m L^2$ and eigenfunctions $\psi_{n_x,n_y}(x,y)=2L^{-1}\sin[n_x\pi(x/L+1/2)]\sin[n_y \pi(y/L+1/2)]$, with $|x|,|y|\leq L/2$. A weighted sum yields the surface density of the gas
\begin{equation}
n_{\mu,T}(x,y)=\sum_{n_x,n_y}{\frac{|\psi_{n_x,n_y}(x,y)|^2}{\exp[(E_{n_x,n_y}-\mu)/\kB T]-1}}.
\label{eq:S2}
\end{equation}
Here, the chemical potential $\mu\leq E_{1,1}$ determines the total particle number, $N=\int{\mathrm{d}x\mathrm{d}y\ n_{\mu,T}(x,y)}$, independently of temperature $T$. Experimentally, we control $\mu$ by varying the excitation level of the dye molecules via the pump laser power, while $T=\SI{300}{\kelvin}$ is fixed by the thermalization of the photons to the dye heat bath, as observed in the broadband spectral photon distributions shown in Fig.~\ref{fig:1}(C) of the main text. Moreover, by performing spectroscopy of the lowest-lying single-particle states of the photon gas, one can validate whether the trapping potential is indeed well-described by a box. Figure~\ref{fig:S2}(E) shows corresponding high-resolution spectra for two boxes of $L=\SI{22(1)}{\micro\meter}$ and $\SI{36(1)}{\micro\meter}$ containing several resonances. The identified resonance frequencies are in good agreement with theory, and for the two boxes $E_{1,1}/2\pi\hbar=\SI{42(3)}{\giga\hertz}$ and $\SI{17(2)}{\giga\hertz}$, respectively, above the low-frequency cavity cutoff $m(c/\tilde n)^2$. A comparison with thermal energy $\kB T/2\pi\hbar\approx \SI{7.5}{\tera\hertz}$ indicates that the eigenenergies– even for small system sizes– can be considered quasi-continuous.

Equation~\eqref{eq:S2} allows for an estimate of the critical particle number $\Nc$ for condensation of the gas in the box. We expand the low-momentum part of the Bose-Einstein distribution $n_k=4\pi\lambda^{-2}/(k^2+k_\mathrm{c}^2)$ with thermal wavelength $\lambda =2\pi\hbar/\sqrt{2\pi m \kB T}$, and find an expression for the correlation length $\xi=k_c^{-1}=\hbar / \sqrt{2m|\mu|}$. Upon increasing $N$, also $\xi$ increases and eventually spans the system size, $\xi(\Nc)=L$. This condition defines a critical particle number
\begin{equation}
\Nc= - \frac{L^2}{\lambda^2} \log\left[ 1- \exp\left(- \frac{\lambda^2}{L^2}\frac{1+2\pi^2}{4\pi}   \right)  \right],
\label{eq:S3}
\end{equation}
which gives $\Nc\approx4500$ for $L=\SI{40}{\micro\meter}$ and $\lambda\approx \SI{1.47}{\micro\meter}$. The log-term gives a critical phase-space density $\mathcal{D}_\mathrm{c}\approx 6.1$, which lies slightly above the measured value of $\mathcal{D}_\mathrm{c,exp}=4.7(5)$, see the inset of Fig.~\ref{fig:2}(C); for $L=\SI{48}{\micro\meter}$, on the other hand, we find $\mathcal{D}_\mathrm{c,exp}=6.2(6)$, in good agreement with the predicted $\mathcal{D}_\mathrm{c}\approx 6.5$. In the limit $L\gg \lambda$, which is fulfilled for our boxes, eq.~\eqref{eq:S3} predicts a scaling $\Nc\sim (L/\lambda)^2 \log(L/\lambda)$, which we confirm experimentally in Fig.~\ref{fig:2}(C) of the main text.

The experimentally achievable box potentials differ from the infinite-depth textbook scenario and exhibit both a finite trap depth and slope width, see Fig.~\ref{fig:S2}. Additionally, the lowest-energy wave functions are altered in the presence of a potential gradient $U_0/L$ if their energy $E_{n_x,n_y}<U_0$ (see section “Compressibility from center-of-mass response”). For theoretical predictions, we employ numerical wave functions obtained by solving the Schrödinger equation with potentials inferred from cuts through the box along $x$ and $y$, respectively. The density distribution is then evaluated using eq.~\eqref{eq:S2}. Since the computation is sensitive to residual detection noise on the measured height profiles of the box nanostructures, we first fit a Gaussian-convolved step function to the surface scan data and use it to model the box potential. To account for the experimentally observed finite trap depth $V_0$, we discard all modes with energy above $1.4 \kB T$. A numerical estimate for the critical particle number, $\Nc\approx 3600$, is obtained in analogy to our measurements in Fig.~\ref{fig:2}(B). To model the photon gas compressibility, we follow the same procedure described above, but superimpose the box with a linear potential $U(x)=U_0 x/L$.

\begin{figure}[t] 
  \centering
  \includegraphics[width=0.6\columnwidth]{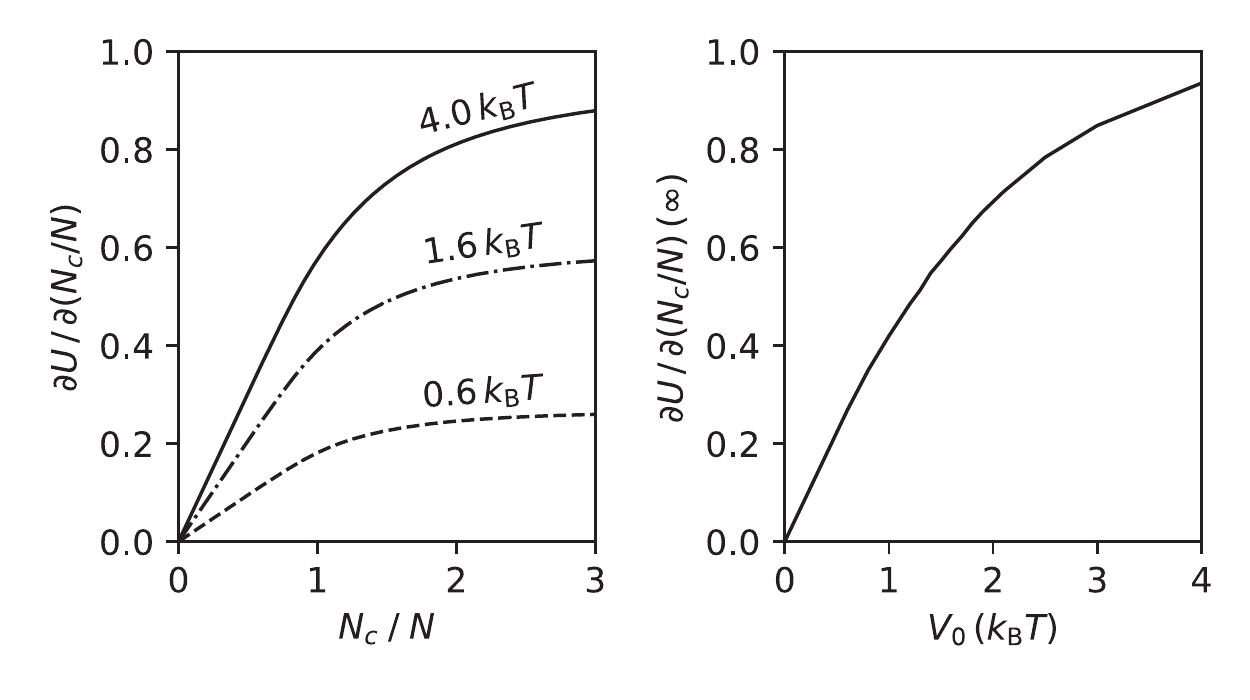}
 \caption{{\bf Effect of trap depth on caloric properties.} Derivative of the internal energy $U$ with respect to $\Nc/N$ for different trap depths $V_0/\kB T = \{0.6,1.6,4.0\}$ (left). In the normal-gas phase, $\Nc/N\gg 1$, the curves saturate at values for the specific heat per particle below the prediction for $V_0\rightarrow\infty$ at a value of 1. The latter is restored as $V_0$ increases (right). For our experiments, $V_0\approx 1.4\kB T$.
}
\label{fig:S3}
\end{figure}

To conclude our discussion of the thermodynamics of the 2D photon gas in the box, we focus on the inverse particle number dependence ($\Nc/N$) of the normalized internal energy $U=\langle E\rangle \Nc/(N^2 \kB T)$ (units of $\kB T$) and its derivative, both shown in Fig.~\ref{fig:3}(B) of the main text. With respect to the cavity low-frequency cutoff $m (c/\tilde{n})^2$, the average transverse energy reads
\begin{equation}
\langle E\rangle = \sum_{n_x,n_y}{\frac{E_{n_x,n_y}-E_{1,1}}{\exp[(E_{n_x,n_y}-\mu)/\kB T]-1}}.
\label{eq:S4}
\end{equation}
The representation $U(\Nc/N)$ gives insight into the effective temperature behavior of the internal energy; the derivative $\partial U/\partial(\Nc/N)$ then provides a qualitative measure of the specific heat. While in a uniform gas we have $\Nc/N=T/\Tc$ with a critical temperature $\Tc$, in the box logarithmic finite-size corrections modify this relation, but $\Nc/N$ remains a purposeful measure for the reduced temperature at the used $L=\SI{40}{\micro\meter}$. In the normal-gas limit, $\Nc/N\gg 1$, one usually expects an internal energy (or specific heat) per particle $U/N=\kB T$ (or $c_V/N=\kB$), owing to the two motional degrees of freedom in the 2D homogeneous system. However, due to our finite $V_0$ the sum in eq.~\eqref{eq:S4} is truncated and both quantities take a value of roughly $0.5$ in their respective units. Numerically, this result is directly obtained from eq.~\eqref{eq:S4} and shown in Fig.~\ref{fig:S3} for increasing $V_0$, for which the infinite-depth prediction is gradually restored. For the analysis of our experimental data, on the other hand, we replace the Bose-Einstein distribution function in eq.~\eqref{eq:S4} with the recorded spectra shown in Fig.~\ref{fig:1}(C) of the main text. Additionally, the sum iterates over energy occupations $n(E)$ (obtained from spectrometer pixels) instead of eigenstates. The photon numbers are calibrated by identifying the spectrum at the critical point, setting it to $\Nc$, and normalizing the total photon number of all other spectra based on the relative integrated signals. The specific heat data is obtained by numerical derivation.

\begin{figure}[t] 
  \centering
  \includegraphics[width=0.95\columnwidth]{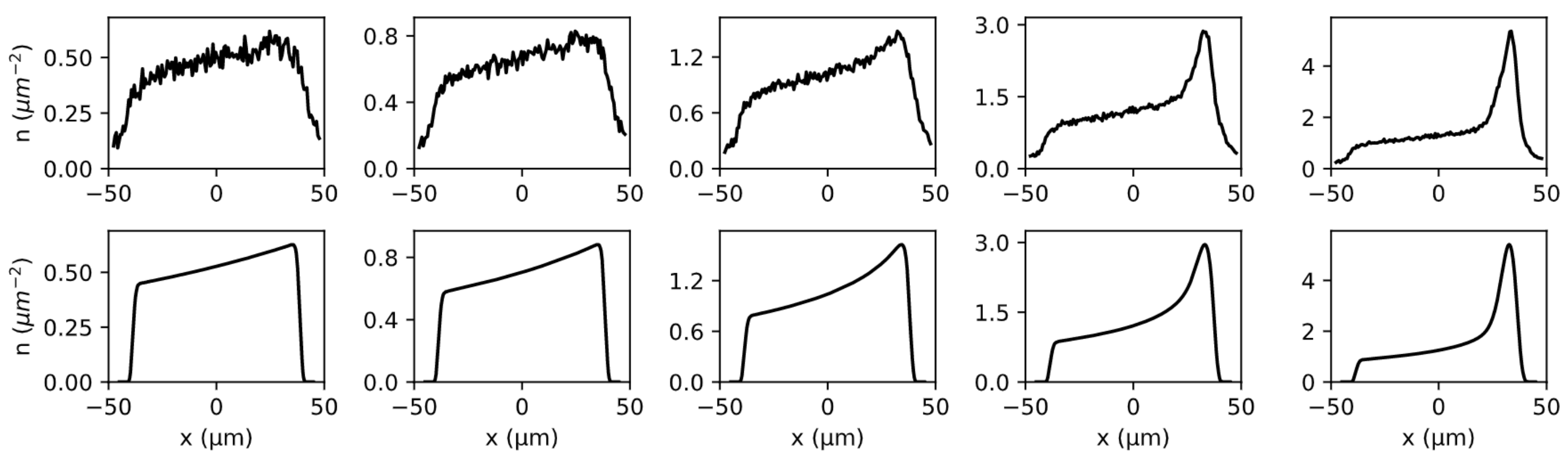}
 \caption{{\bf Density distributions of the photon gas in a tilted box.} Exemplary line profiles of the experimental (top row) and theoretical (bottom) density distributions in a box of size $L=\SI{80}{\micro\meter}$ at photon numbers $N\approx\{3000, 4500, 6800, 8500\}$ (from left to right). The tilt potential $U_0/\kB\approx\SI{43}{\kelvin}$ induces a visible density gradient along the $x$ direction and displaces the cloud’s center-of-mass, which is more pronounced for larger $N$.
}
\label{fig:S4}
\end{figure}

\subsection{Compressibility from center-of-mass response}
The isothermal compressibility $\kappa_T = -V^{-1}(\partial V/\partial P)_T = n^{-2}(\partial n/\partial \mu)_T = [\exp(n\lambda^2)-1]/(\kB T n^2 \lambda^2)$ of the infinite 2D Bose gas, as calculated from the equation of state $n(\mu) = -\lambda^{-2}\log[1-\exp(\mu/\kB T)]$, diverges in the quantum degenerate regime towards large densities $n$, as indicated in Fig.~\ref{fig:4}(B) (dashed line) of the main text. Experimentally, this pathological feature of the infinite 2D homogeneous gas cannot be investigated. A finite-size system, however, enables the study of the highly-compressible ideal gas, where the energy cost to increase the particle density gradually vanishes, $\partial\mu/\partial n\rightarrow 0$ and the volume of the gas (here: surface area) can be reduced by applying infinitesimal pressures. Experimentally, we probe the compressibility of our photon gases by squashing the photons in the box. A linear gradient potential $U(x)=U_0 x/L$ is applied along $x$ by tilting one of the cavity mirrors (see section “Experimental scheme”), which results in a spatially uniform force $F_0=U_0/L$ acting on the photons and displacing the cloud's center-of-mass
\begin{equation}
\langle x\rangle = L \int_{-L/2}^{L/2}{\mathrm{d}x\ x \frac{n_{\mu,T}(x)}{N}}
\label{eq:S5}
\end{equation}
with surface density $n_{\mu,T}(x)$ along $x$ at chemical potential $\mu$ and temperature $T$ and photon number $N=L\int{\mathrm{d}x\ n_{\mu,T}(x)}$. Figure~\ref{fig:S4} gives experimental and theoretical density line profiles in a tilted box. Within certain limits, $\langle x\rangle$ gives a direct measure of $\kappa_T$, as we derive below.

\begin{figure}[t] 
  \centering
  \includegraphics{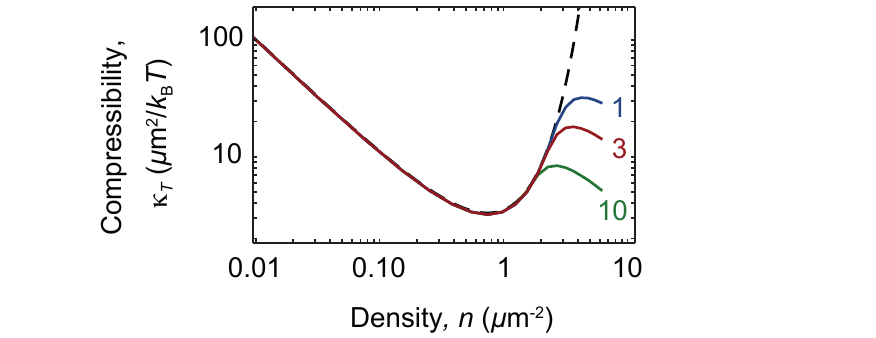}
 \caption{{\bf Validity of the compressibility extraction.} Numerical study of the compressibility $\kappa_T$ extracted from the center-of-mass in tilted boxes with potentials $U_0/\kB =\{1,3,10\}\SI{}{\kelvin}$ (blue, red, green lines) versus density $n$, along with the theory prediction for a 2D ideal Bose gas in a non-tilted box (solid black) and for the infinite system (dashed). The extraction in the degenerate region at densities $n\gtrsim \SI{1}{\micro\meter}^{-2}$ becomes more accurate for smaller tilts $U_0$.
}
\label{fig:S5}
\end{figure}

Our approach is based on the local density approximation, which connects $\mu$ of a spatially inhomogeneous system with the chemical potential of the homogeneous system $\mu_0$ via the position-dependent potential energy; for the case of a linear gradient potential, $\mu(x) = \mu_0 - U_0 x/L$. We perform a variable transformation with $x = (\mu_0-\mu)L/U_0$ and $\mathrm{d}x = -L/U_0 \mathrm{d}\mu$ to eliminate the position dependence and express the density as $n(\mu)$. From eq.~\eqref{eq:S5},
\begin{equation}
\langle x\rangle = \frac{L^3}{N U_0^2} \int_{\mu_0+U_0/2}^{\mu_0-U_0/2}{\mathrm{d}\mu (\mu-\mu_0)n(\mu)} \simeq \frac{L^3}{N U_0^2} \int_{\mu_0+U_0/2}^{\mu_0-U_0/2}{\mathrm{d}\mu(\mu-\mu_0) \left[ n(\mu_0) + (\mu-\mu_0) \left. \frac{\partial n}{\partial \mu} \right|_{\mu_0} \right] },
\label{eq:S6}
\end{equation}
where we have expanded $n(\mu)$ to first order, indicating how the compressibility $\kappa_T = n^{-2}(\partial n/\partial \mu)_T$ enters the center-of-mass. Integration yields the relation used to extract $\kappa_T$:
\begin{equation}
\frac{\langle x\rangle}{L}= - \frac{U_0}{12}\frac{L^2}{N} \left.\frac{\partial n}{\partial\mu}\right|_{\mu_0} = -\frac{\kappa_T n}{12} U_0
\label{eq:S7}
\end{equation}
Note that the linearization in eq.~\eqref{eq:S6} is valid only if the potential amplitude $U_0$ remains within certain limits, which differ for the normal ($U_0\ll \sqrt{40}\kB T$) and the condensed ($U_0\ll \sqrt{20}\mu$) region; for the latter case where $\mu\rightarrow 0$, this requires very small tilts $U_0$. While the general applicability of the method remains valid, it requires the detection of small displacements of the photon gas at weak tilts. Figure~\ref{fig:S5} shows a numerical calculation of the $\kappa_T$ extraction using the center-of-mass method for three values of $U_0$ (solid lines), indicating that small values of $U_0$ indeed improve the agreement with the theory prediction for the 2D Bose gas in an untilted box. Our experimental data in Fig.~\ref{fig:4}(A) highlights the change in sensitivity, on the one hand confirming linearity $\langle x\rangle (U_0)$ over a wide range of values of $U_0$ in the normal phase, on the other hand revealing a nonlinear response for the degenerate gas. For the $\kappa_T$ extraction, in all cases we analyze the (narrow) linear response region only.

\begin{figure}[t] 
  \centering
  \includegraphics[width=0.85\columnwidth]{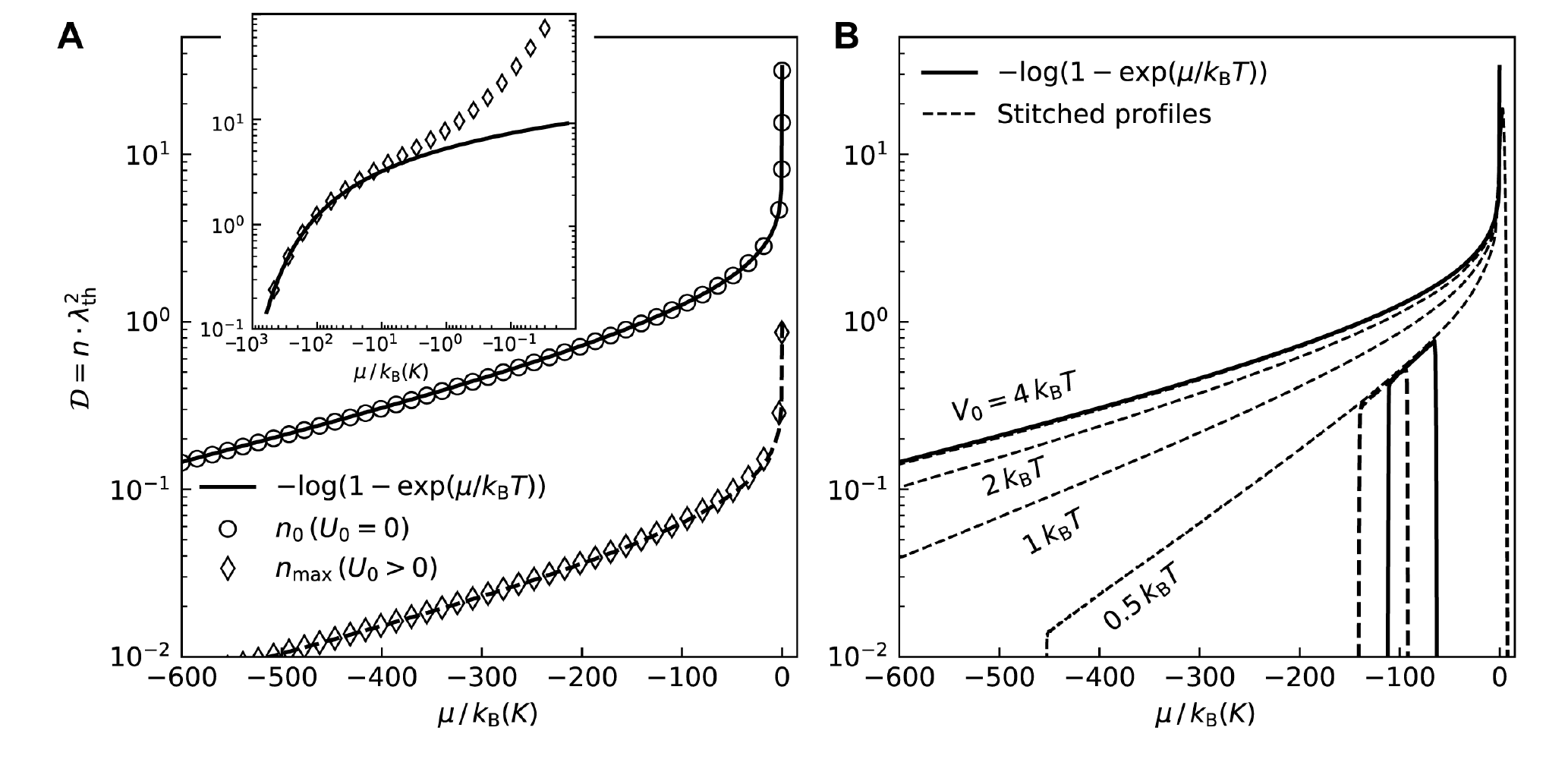}
 \caption{{\bf Equation of state reconstruction by density profile stitching.} ({\bf A}) The EOS of the infinite 2D homogeneous ideal Bose gas (solid line) can be reconstructed from the central density of a Bose gas in a finite-size box (circles) in the non-degenerate regime. The method works also when extracting the maximum density of the gas in a box superimposed with a gradient potential $U_0 x/L$ (diamonds); here, both infinite-system EOS (dashed) and data were shifted down. Inset: In the quantum degenerate regime, the reconstruction loses its validity. ({\bf B}) Illustration of the numeric stitching process, which is exact for infinitely deep boxes (numerically confirmed for $V_0>4\kB T$, deviations at $\mu=0$ due to the spatial extent of the ground mode) but deviates for smaller $V_0$. Two examples of calculated density profiles (solid and dashed) used for the stitching are shown for the $0.5\kB T$ curve.
}
\label{fig:S6}
\end{figure}

\subsection{Reconstruction of the equation of state}
The equation of state (EOS) in the form $n=f(\mu,T)$ is a monotonous function, see Fig.~\ref{fig:S6}(A); hence the linearly increasing photon density profiles $n(x)$, see Fig.~\ref{fig:S4}, observed in the presence of the linear potential allow for a reconstruction of the EOS. In the experiment, $\mu(x,N)$ is controlled both by varying the total photon number $N$ or by applying a potential such that $\mu(x)\rightarrow \mu_0 - U_0 x/L$ (local density approximation). A single density profile partially yields the EOS over an energy range of width $\mu=U_0$. To reconstruct $n(\mu)$ over larger ranges of $\mu$, multiple profiles with partially overlapping densities, say $n_1(x)$ and $n_2(x)$, but different $N$ are recorded at fixed $U_0$ and stitched together. A relative shift by $\Delta\mu$, required to match $n_1(\mu)= n_2(\mu + \Delta\mu)$, merges the profiles into a continuous trace, see Fig.~\ref{fig:S6}(B). Since experimental density profiles are subject to noise, we use linear fits (excluding boundary of the box) as a reference for the stitching procedure, which is carried out for $\sim$300 traces. For $\mu/\kB > -\SI{50}{\kelvin}$, where the density profiles are observed to become nonlinear, a corresponding set of profiles containing the finite-size condensate peak is stitched manually.

We numerically find that due to finite trap depth $V_0$, the slope of $n(\mu)$ increases, see Fig.~\ref{fig:S6}(B). Note that this implies a larger compressibility $\kappa_T = n^{-2}(\partial n/\partial\mu)_T$, as observed in Fig.~\ref{fig:4}(B) of the main text. By calculating the surface density of the photon gas in a finite-depth box and comparing it to its infinite-depth counterpart, we analytically find from Boltzmann statistics that rescaling $\mu\rightarrow\mu/[1-\exp(-V_0/\kB T)]$ in the infinite-system EOS $n(\mu)$ reproduces the finite-depth result (solid line in Fig.~\ref{fig:4}(C) of the main text).

\end{document}